\documentclass[]{spie}  

\usepackage{amsmath,amsfonts,amssymb}
\usepackage{graphicx}
\usepackage[colorlinks=true, allcolors=blue]{hyperref}
\usepackage{siunitx}
\usepackage{appendix}
\usepackage{subcaption}

\title{Polarization aberration modeling of internal occulters for coronagraphs}

\author[a,*]{Emory L. Jenkins}
\author[b]{A J Eldorado Riggs}
\author[c,*]{Ewan S. Douglas}
\author[c]{Ramya M. Anche}
\author[b]{Dylan M. McKeithen}
\author[b]{Stuart B. Shaklan}
\affil[a]{James C. Wyant College of Optical Sciences, University of Arizona, Tucson, AZ, 85721}
\affil[b]{Jet Propulsion Laboratory, California Institute of Technology, Pasadena, CA 91109}
\affil[c]{Department of Astronomy and Steward Observatory, University of Arizona, Tucson, AZ,
85721}

\authorinfo{Send correspondence to E.L.J: emoryljenkins@arizona.edu \\
\copyright 2025. All rights reserved.}

\begin{document} 
\maketitle

\begin{abstract}
The Habitable Worlds Observatory (HWO) is a flagship mission concept proposing to characterize earth-like exoplanets at high contrast with a coronagraph instrument. 
The most in-depth, validated contrast error budgets made to date have been at the $10^{-9}$ planet-to-star contrast levels for the Roman Space Telescope.
To obtain the raw contrast levels of $\leq 10^{-10}$ contrast needed for HWO, more modeling is needed of the vector diffraction effects within a coronagraph instrument. 
Several coronagraph architectures are based on the Lyot coronagraph, which utilizes a small occulting spot to block most of the on-axis starlight at a focal plane. 
In this paper, we present a study of the polarization aberrations generated by internal occulting spots.
First, we use finite-difference time-domain (FDTD) modeling to produce the complex transmission of the occulter in each polarization state, and then we separately run those through a closed-loop wavefront control model in FALCO\cite{rigss2018falco} to determine the degradation in achievable contrast. 
We perform sweeps of such parameters as spot diameter, spot thickness, and angle of incidence to inform the HWO instrument design. 
\end{abstract}


\section{INTRODUCTION}
\label{sec:intro}

Next-generation space telescopes such as the Habitable Worlds Observatory (HWO) seek to directly image terrestrial exoplanets in order to detect their compositions and signatures of life. 
Due to such a planet's size and proximity to its star, the instrument will need to suppress the starlight by $10^{-10}$ or better in visible light.\cite{Snellen2022whitepaper} 
The current generation coronagraph instrument achieves $10^{-9}$ suppression at best,\cite{kasdin2020nancy,bailey2023nancy} making the HWO's goals a significant step forward from the state of the art.
When driving contrast towards the $10^{-10}$ level, extreme wavefront control is necessary to both suppress aberrations and null any starlight that leaks into the focal plane via techniques such as electric field conjugation (EFC).\cite{giveon2007efc}
However, active wavefront control via deformable mirrors (DMs) cannot, in general, correct for polarization aberrations.\cite{Breckinridge2015polabs} 
There is a concern that polarization aberrations will represent a significant fraction of residual starlight reaching the science focal plane at $10^{-10}$ contrast. 
To ensure adequate performance, there is a need to set a contrast budget for each component of a coronagraph instrument that may contribute to leaking starlight.

In this manuscript, we consider the small focal-plane occulting spots used in Lyot coronagraphs to block on-axis signals.
These occulters are usually on the order of 100 \si{\micro \meter} in diameter and made by depositing a $\sim 100-200$ \si{\nano \meter} of metal film on a transmissive substrate, such that any remaining on-axis starlight is rejected.
Since an occulter of this type is somewhat small in diameter relative to visible wavelengths ($\sim 20-200\lambda$), the fractional area within one wavelength of the edge is large enough to raise concern with regard to polarization effects.
This is because the edges of the metal spots will have different responses to orthogonal polarization states.

With limited space available on an observatory such as HWO, a $2\times$ faster beam at the occulter would be favorable from an optical design standpoint.
However, a $\sim 50$ \si{\micro \meter} occulter may exacerbate starlight leakage due to vector diffraction.
Vector vortex coronagraph (VVC) masks often have $\sim 10-20$ \si{\micro \meter} diameter spots deposited over their singularities to mask out defects.
Modeling VVCs with such dotmasks is outside the scope of this manuscript, but $10$ \si{\micro \meter} and $20$ \si{\micro \meter} occulters are shown to introduce problematic polarization effects in Lyot coronagraphs.

In order to model the polarization interactions of occulting spots, we use the finite-difference time-domain (FDTD)\cite{Yee1966fdtd} method of electromagnetic simulation to extract a full Jones matrix transmission coefficient.
FDTD simulations were made for occulters with a selection of materials, optical densities, diameters, and edge profiles to analyze the polarization effects.
These Jones matrices were added to simple DM-apodized Lyot coronagraph models implemented in the FALCO software package to simulate starlight leakage due the occulters.
Using these results, we set contrast budget terms for the various configurations of the occulting spots.

\section{FDTD Modelling of Occulters}
The FDTD method allows us to model the full three-dimensional, vectorized interaction of light with the occulters. 
At its core, the FDTD model numerically steps forward through time a system of differential equations.
In this case, Maxwell's equations of the electric and magnetic fields are the relevant governing equations.
For FDTD modeling, we use the MIT Electromagnetic Equation Propagator (Meep), a robust, open source FDTD package with a Python interface.\cite{oskooi2010meep}

\begin{figure}
    \centering
    \includegraphics[width=0.5\linewidth]{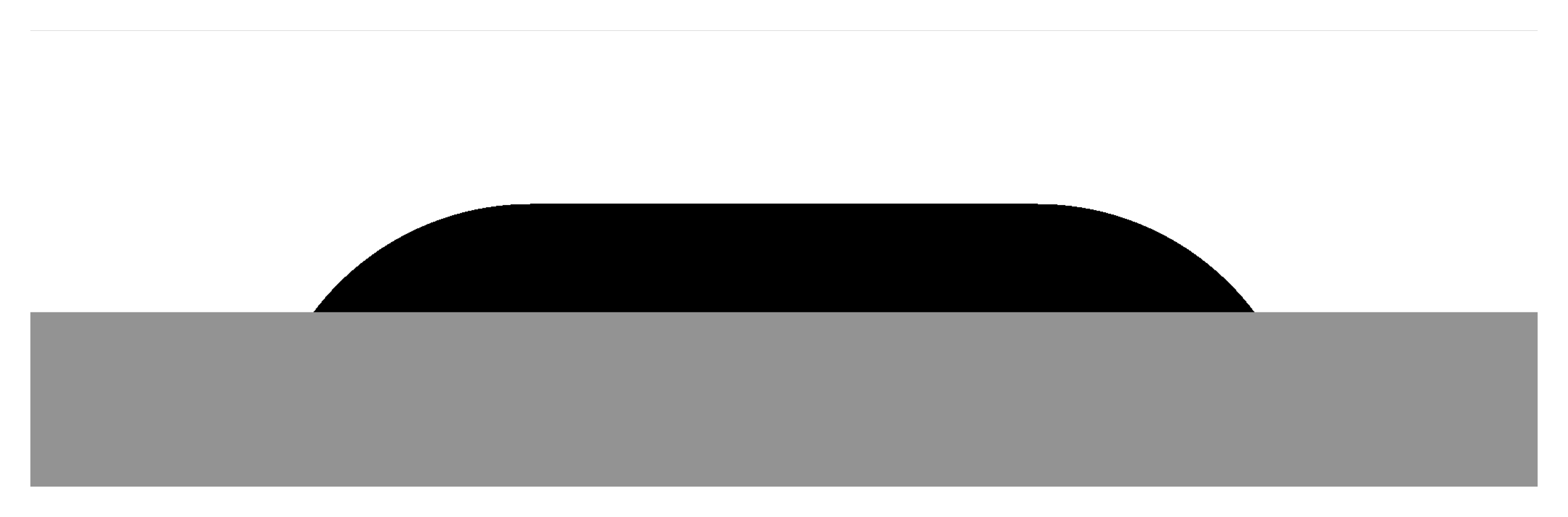}
    \caption{Cross-sectional view of occulting spot with radiused edges (black) on a substrate (grey).}
    \label{fig:cross_sec}
\end{figure}

Occulters are usually made by depositing a metal dot via liftoff photolithography on a transmissive substrate that is often coated with an antireflection stack, though we choose to ignore any coatings for simplicity.
The initial occulter model is a metal cylinder with its base in contact with a slab of fused silica.
In reality, an occulting spot will have smooth edges due to the deposition process, rather than the sharp right angle of the top of a cylinder.
To approximate these smooth edges, we choose to make the cross-sectional height at the edge of the spot decrease as a circular arc.
The width of the tapered region is a controlled parameter in the simulations, and the radius of the arc is chosen such that the top surface of the spot is tangential to the arc, as shown in Figure \ref{fig:cross_sec}.

\begin{figure}
    \centering
    \includegraphics[width=0.5\linewidth]{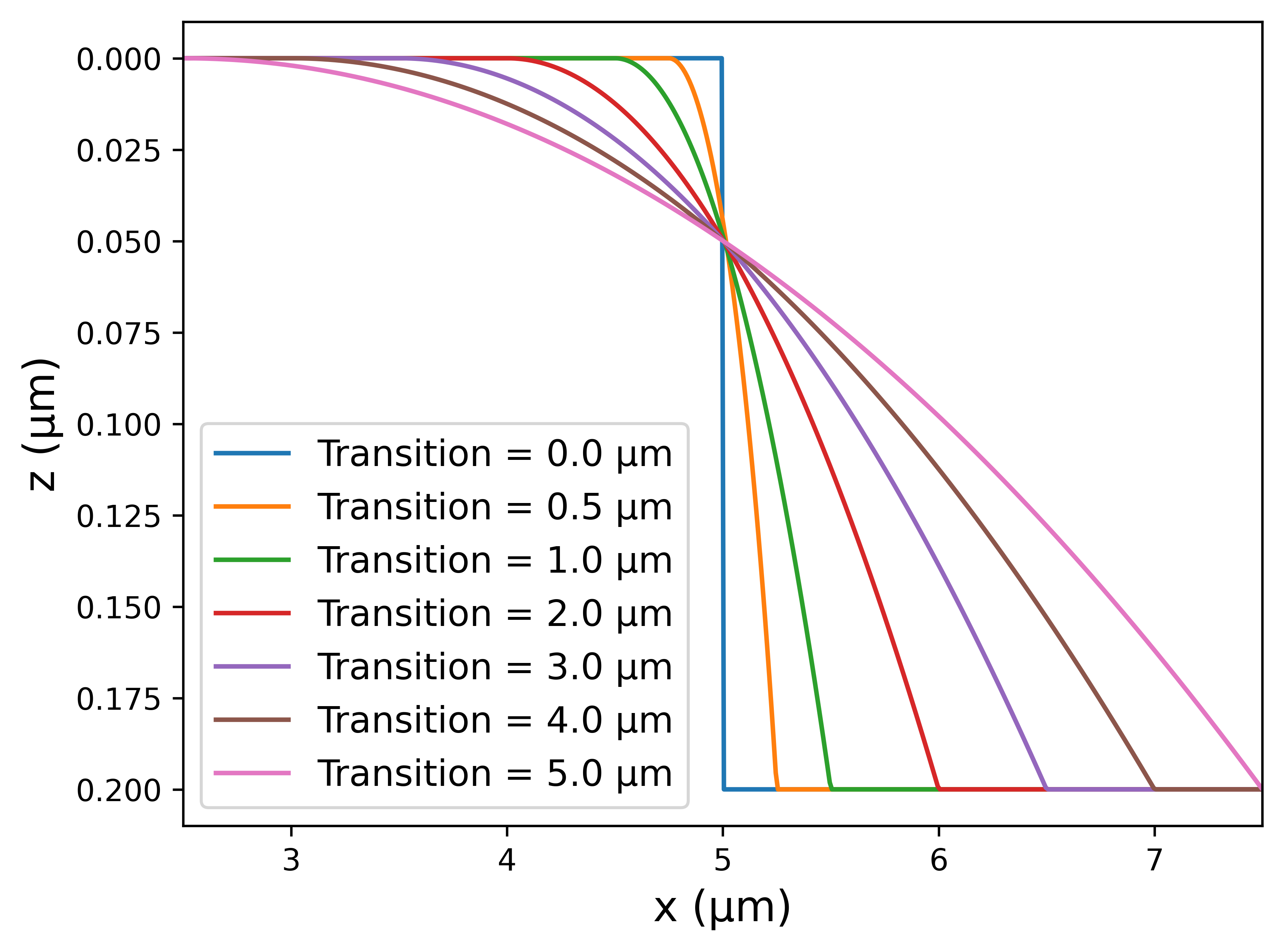}
    \caption{Edge profiles of occulters for various widths of transition regions for a $10$ \si{\micro \meter} occulter. Note how the center of the transition region remains at the nominal radius of the spot.}
    \label{fig:transition curves}
\end{figure}

The FDTD simulation volume is made to be three times the occulting spot diameter in the x and y directions, and $600$ \si{\nano \meter} greater than the spot thickness in the z direction.
This leaves enough room to inject a Gaussian linewidth planewave just before the occulting spot and for surface plasmons at the rear surface of the spot to decay.
Since we are interested in calculating the focal plane mask's complex transmission of the electric field $J_{fpm}$, we need to extract 6 fields from FDTD simulation. 
$E_h^{in}$ and $E_v^{in}$ are simply the injected electric fields from separate FDTD simulations without an occulting spot, measured within the substrate for normalization. 
Here, $h$ and $v$ refer to the vertical and horizontal linear polarization basis.
$E_{hh}^{out}$, $E_{hv}^{out}$, $E_{vh}^{out}$, and $E_{vv}^{out}$ are the electric fields measured at the same location within the substrate, but with the occulting spot included in the simulation. The $hh$, $hv$, $vh$, and $vv$ designations refer to the output polarization state followed by the input state.
\begin{equation}
    J_{fpm} =
    \begin{pmatrix}
        j_{hh} & j_{hv} \\
        j_{vh} & j_{vv}
    \end{pmatrix} =
    \begin{pmatrix}
        E_{hh}^{out}/E_h^{in} & E_{hv}^{out}/E_v^{in} \\
        E_{vh}^{out}/E_h^{in} & E_{vv}^{out}/E_v^{in}
    \end{pmatrix}
    \label{eq:jones}
\end{equation}

\subsection{Two-Dimensional FDTD Simulation}
With the exception of angle of incidence, the other parameters investigated do not break the rotational symmetry of the structures.
When rotational symmetry is present, all of the relevant interactions between fields and structures may be captured via two-dimensional (2-D) FDTD simulations rather than three-dimensional (3-D). 
The advantage is an n-fold reduction in compute time and memory use, where n is the number of samples in the collapsed dimension.

The 2-D simulation domain is a cross-sectional slice of the occulter.
Since the spot has radial symmetry, its response to $v$ polarization along the vertical direction is identical to its response to $h$ polarization along the horizontal direction, and vice-versa.
Therefore, in order to get all six terms of Equation \ref{eq:jones} from a single cross-section, it makes sense to move to the $s$, $p$ basis.
$E_s^{in}$, $E_p^{in}$, $E_{ss}^{out}$, and $E_{pp}^{out}$ are one-dimensional arrays, and therefore are interpolated radially in to 2-D arrays. 
The transformation to the $v$, $h$ basis is given in Equations \ref{eq:reconstruction}, where $\theta$ is the azimuthal coordinate.

\begin{subequations}
    \begin{equation}
        j_{hh} = t_{ss}\sin^2{\theta} + t_{pp}\cos^2{\theta}
    \end{equation}
    \begin{equation}
        j_{vv} = t_{ss}\cos^2{\theta} + t_{pp}\sin^2{\theta}
    \end{equation}
    \begin{equation}
        j_{vh} = (t_{ss}-t_{pp})\sin{\theta}\cos{\theta}
    \end{equation}
    \begin{equation}
        j_{hv} = (t_{ss}-t_{pp})\sin{\theta}\cos{\theta}
    \end{equation}
    \label{eq:reconstruction}
\end{subequations}

\section{Results of FDTD Simulations}
Two-dimensional FDTD simulations allowed a range of parameters to be probed at a resolution of $200$ points per micron due to the relatively light computational load.
This resolution was chosen since it agrees with simulations at $400$ points per micron to $5$ decimals.
The parameters investigated include the metal used, the optical density of the spot, the width of the transition region, and the diameter of the spot.
We consider aluminum and chromium for the spots as these have been used in high-contrast testbeds in the past. Chromium adheres well directly to glass, and aluminum provides high reflectance and higher optical density for a given thickness. We tested optical densities of $5$ and $6$ for each, resulting in thicknesses of $76$ \si{\nano \meter} and $91$ \si{\nano \meter} for Al, and $159$ \si{\nano \meter} and $191$ \si{\nano \meter} for Cr.

\begin{figure}
    \centering
    \includegraphics[width=0.7\linewidth]{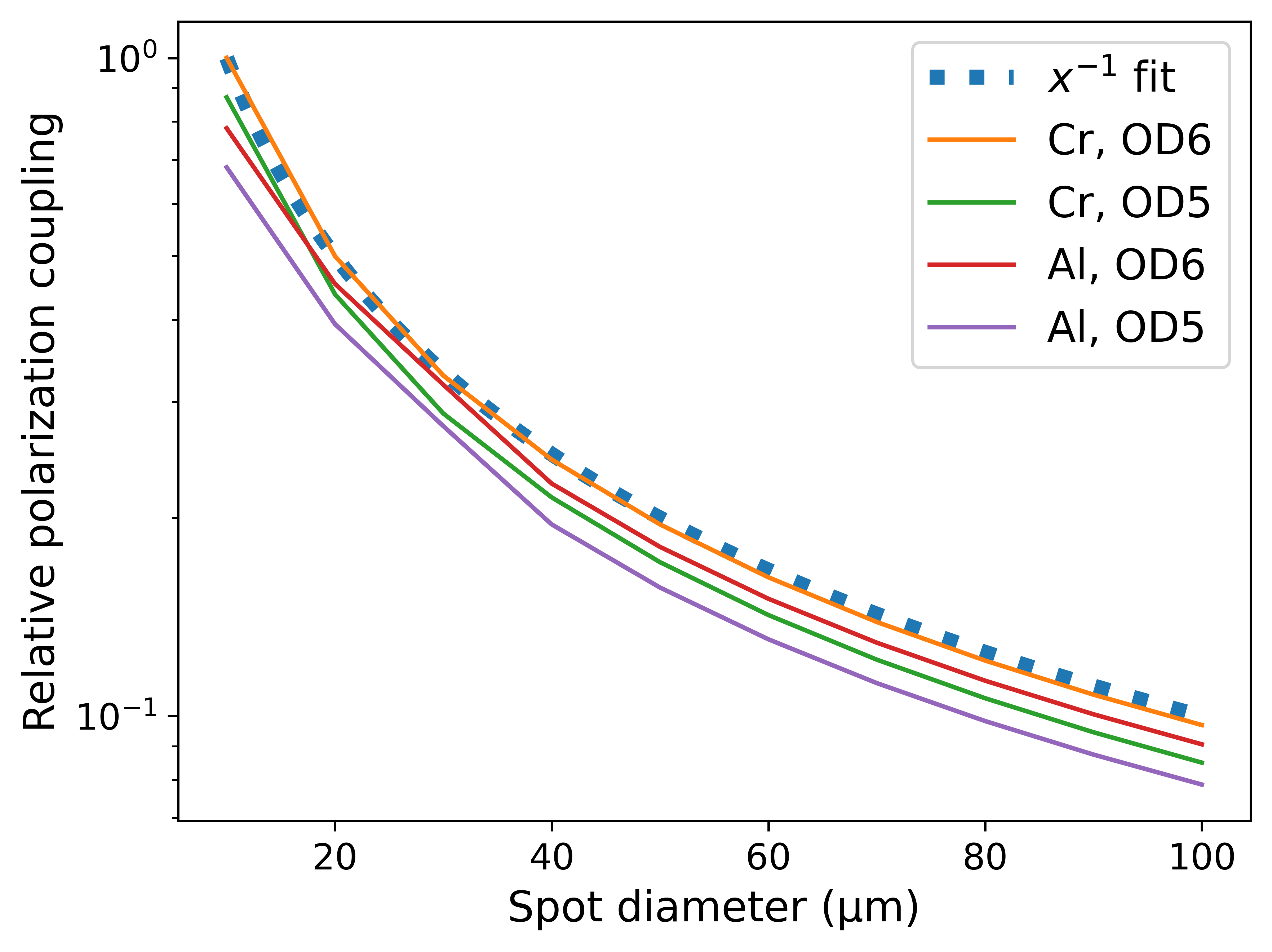}
    \caption{Fractional coupled energy between polarization states as a function of spot diameter, with cylindrical spots rather than sloped edges at $575$ \si{\nano \meter}. As predicted, the coupled energy follows the same $1/D$ scaling as the circumference to area ratio.}
    \label{fig:coupling_diam}
\end{figure}

\begin{figure}
    \centering
    \includegraphics[width=0.7\linewidth]{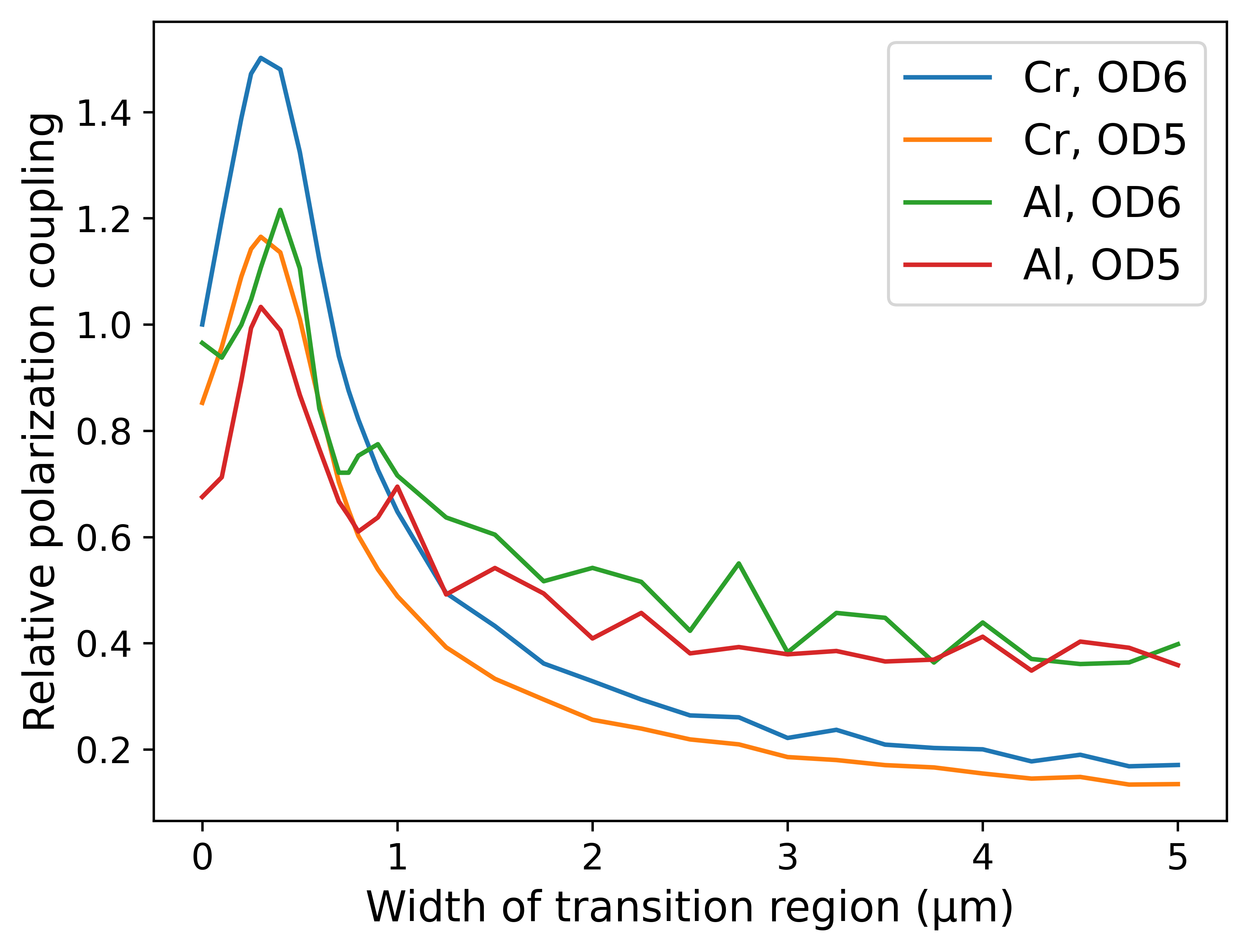}
    \caption{Fractional coupled energy between polarization states as a function of transition region width for $10$ \si{\micro \meter} diameter spots at $575$ \si{\nano \meter}.}
    \label{fig:coupling_transition}
\end{figure}

By taking the $j_{hv}$ or $j_{vh}$ terms of the calculated Jones matrix for each configuration, we can form a relationship between the cross-coupled energy and the probed parameters.
As a function of the occulting spot diameter, shown in Figure \ref{fig:coupling_diam}, the cross-coupled energy decreases with increasing diameter, following a $1/D$ scaling.
This is expected given the area versus circumference of a circle, which scale with $D^2$ and $D$, respectively, and is a trend regardless of the edge profile of the spot.
As a function of edge transition width, shown in Figure \ref{fig:coupling_transition}, the cross-coupled energy increases by upwards of $50\%$ as transition width increases from $0$ \si{\micro \meter} to $\sim 0.4$ \si{\micro \meter}.
Beyond this, the cross-coupling falls significantly for both Cr and Al, though the effect is much stronger for Cr, which reaches $<20\%$ its $0$ \si{\micro \meter} transition coupling by $5$ \si{\micro \meter} transition width.
At $5$ \si{\micro \meter} transition width, both optical densities for Al cross-couple roughly double the energy of the Cr, despite having lower coupling with $0$ \si{\micro \meter} transition.

Using three-dimensional FDTD, simulations for a $10$ \si{\micro \meter} cylindrical (transition region of $0$ width) OD $6.0$ Al occulter were performed for angles of incidence $0^{\circ}$, $5^{\circ}$, and $10^{\circ}$.
This was the largest diameter that we could simulate at 100 points per micron on a supercomputing cluster with four terabytes of RAM.
In 2-D simulations, $100$ points per micron resolution yields results that agree with $200$ points per micron to $4$ decimals, so we consider the simulation converged.

For these simulations, the wavevector lies in the $X-Z$ plane.
That is, the $h$ state is $p$ polarization and the $v$ state is $s$ polarization.
For $5^{\circ}$ angle of incidence (AOI), $h$ to $v$ coupling increases to $1.01$, while the $v$ to $h$ coupling decreases to $0.976$ of the corresponding values for the normal incidence case.
This gap widens at $10^{\circ}$ AOI, with $h$ to $v$ and $v$ to $h$ coupling reaching $1.11$ and $0.947$, respectively.
Despite AOI reducing coupling from $v$ to $h$, the $h$ to $v$ coupling outweighs. Therefore, with an unpolarized beam the total coupling or incoherent fraction of the beam is increased with AOI.

\begin{figure}
    \centering
    \includegraphics[width=0.9\linewidth]{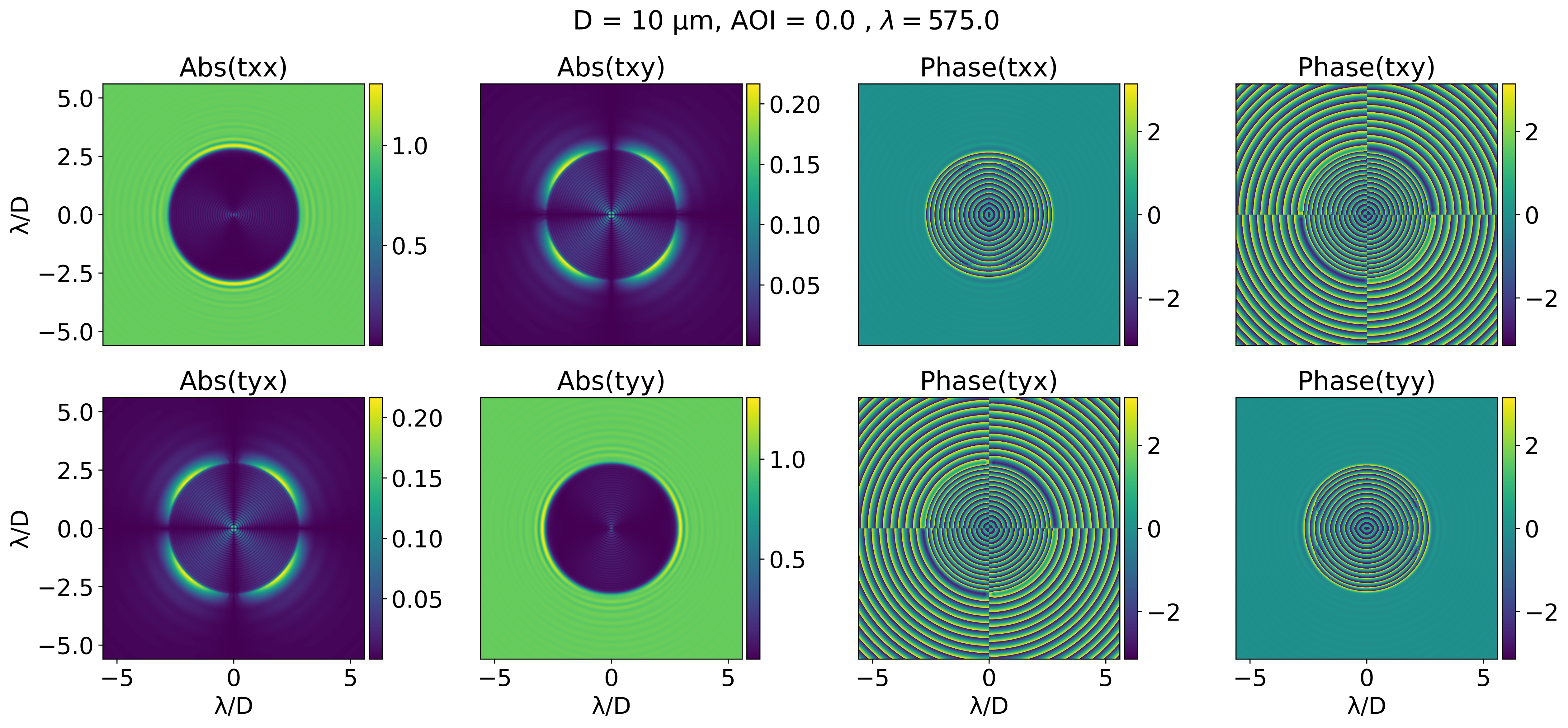}
    \caption{Jones transmission matrix for a $10$ \si{\micro \meter} diameter cylindrical aluminum spot at normal incidence and $575$ \si{\nano \meter}.}
    \label{fig:normal_jones}
\end{figure}

\begin{figure}
    \centering
    \includegraphics[width=0.9\linewidth]{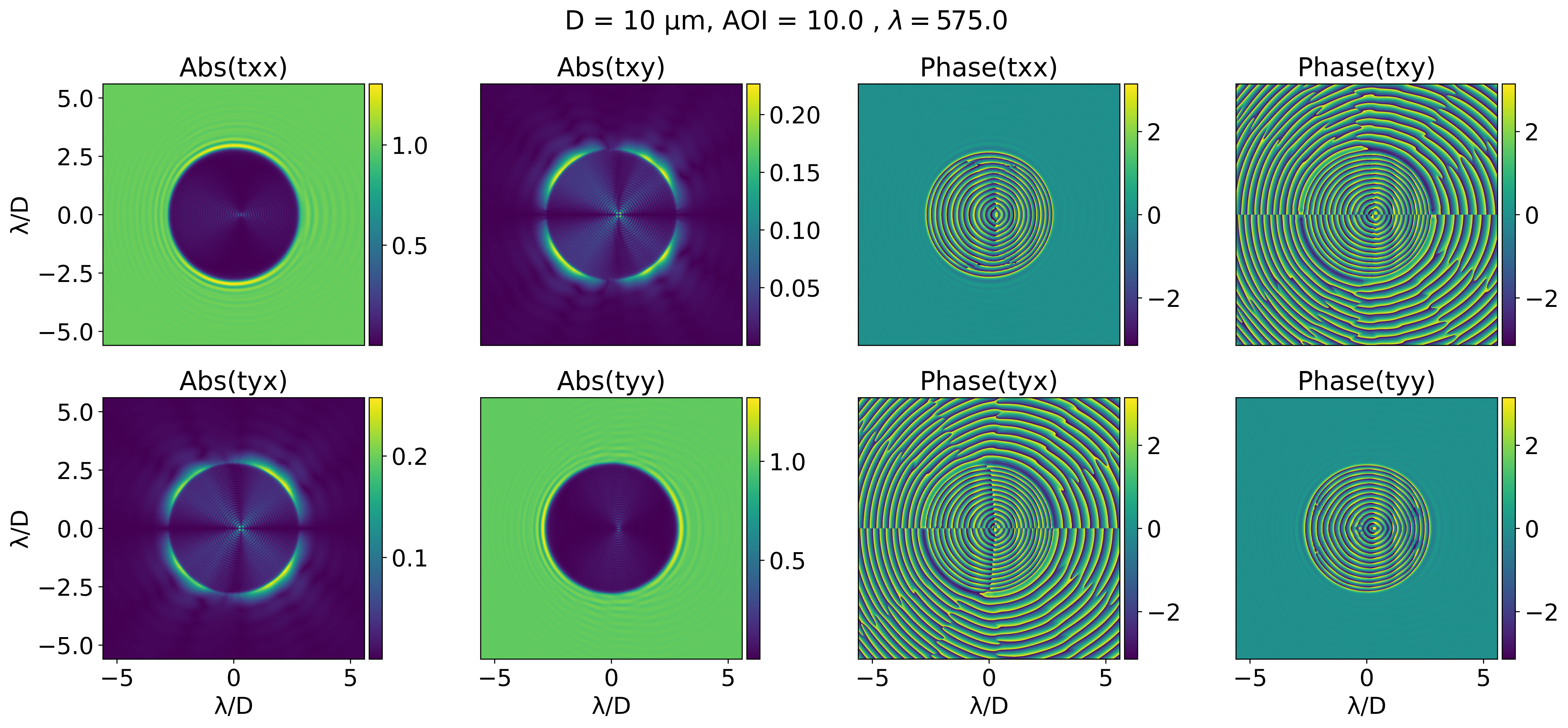}
    \caption{Jones transmission matrix for a $10$ \si{\micro \meter} diameter cylindrical aluminum spot at $10^{\circ}$ incidence and $575$ \si{\nano \meter}.}
    \label{fig:aoi_jones}
\end{figure}

\section{Coronagraphic Results}
\label{falco-results}
Coronagraph simulations were performed in FALCO\cite{rigss2018falco} for an aberration-free, DM-apodized Lyot coronagraph using two $48\times 48$ actuator Xinetics DMs, with $46.3$ actuators across the beam.
Using electric field conjugation (EFC), the DMs first operate on a lower-resolution, scalar diffraction model of the occulter.
Once this scalar dark hole has been dug, the polarization aberrations induced by the occulter are included, and the dark hole is tuned up with several more EFC iterations operating on the mean electric field of the four polarization states as that is what pairwise probing would sense. Finally, the contrast is calculated at this final setting.
Contrast is defined here as the normalized intensity (NI) in the focal plane, where the normalization value is the maximum intensity recorded in the focal plane without an occulter and with a flat wavefront (no DM).
For all configurations, regardless of the occulting spot diameter in microns, the occulter radially subtends $2.8 \lambda/D$ in angle space.
Therefore, the inner working angle (IWA) of the dark hole region is approximately $3.0 \lambda/D$.

\begin{figure}[ht]
    \centering
    \begin{subfigure}{0.4\linewidth}
        \centering
        \includegraphics[width=\linewidth]{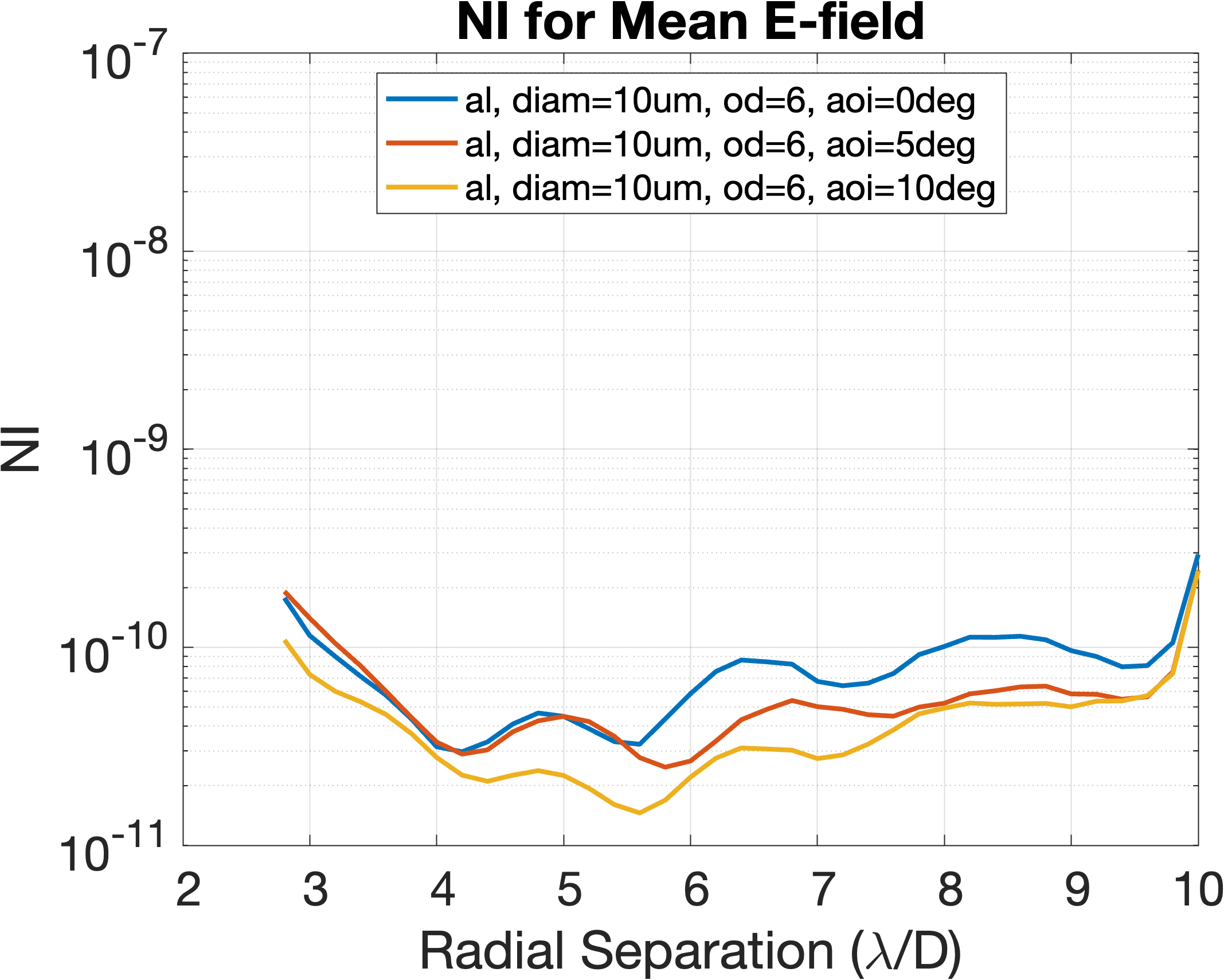}
        \caption{}
        \label{fig:falco_aoi_mean_efield}
    \end{subfigure}
    \begin{subfigure}{0.4\linewidth}
        \centering
        \includegraphics[width=\linewidth]{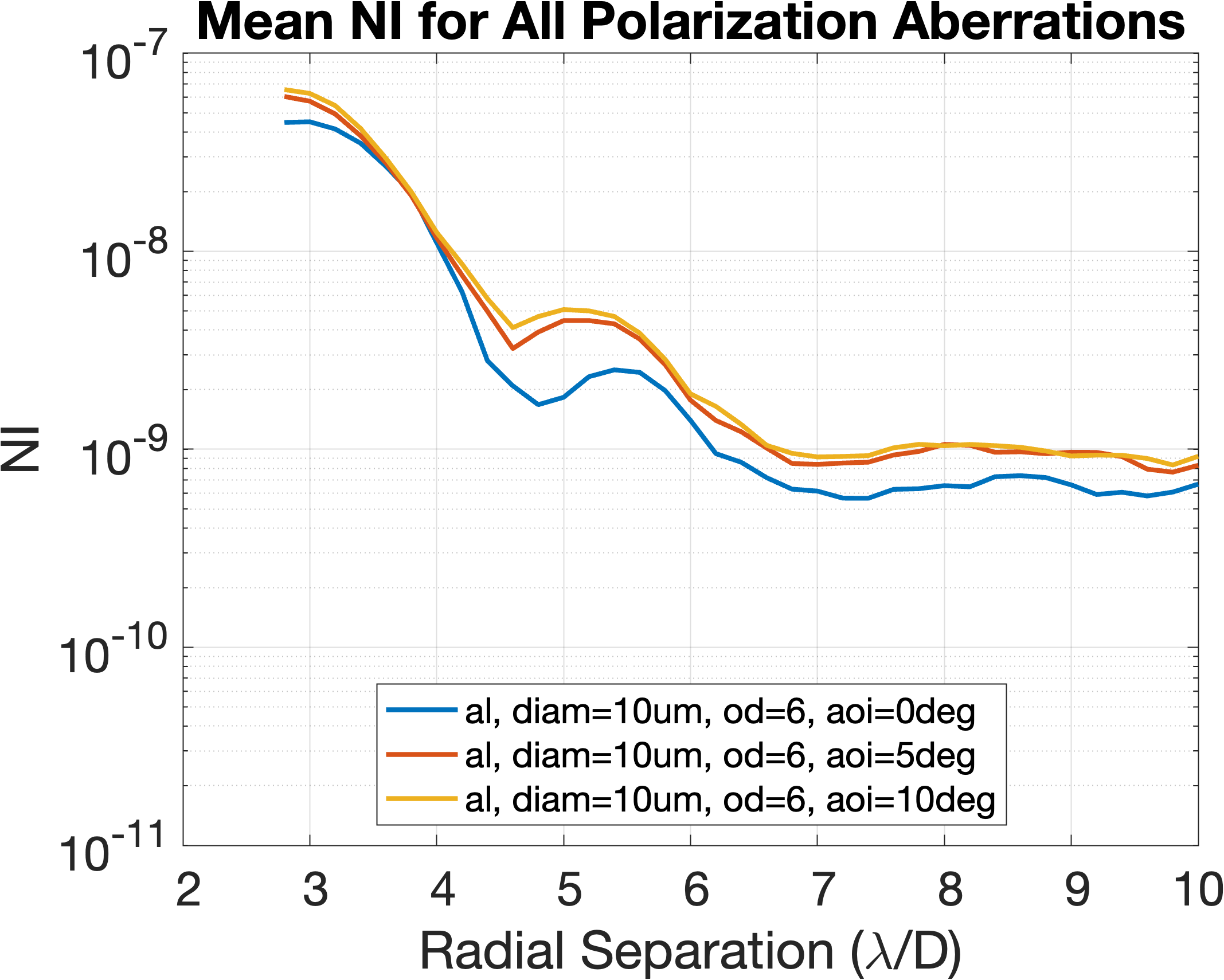}
        \caption{}
        \label{fig:falco_aoi_mean_polab}
    \end{subfigure}
    \begin{subfigure}{0.4\linewidth}
        \centering
        \includegraphics[width=\linewidth]{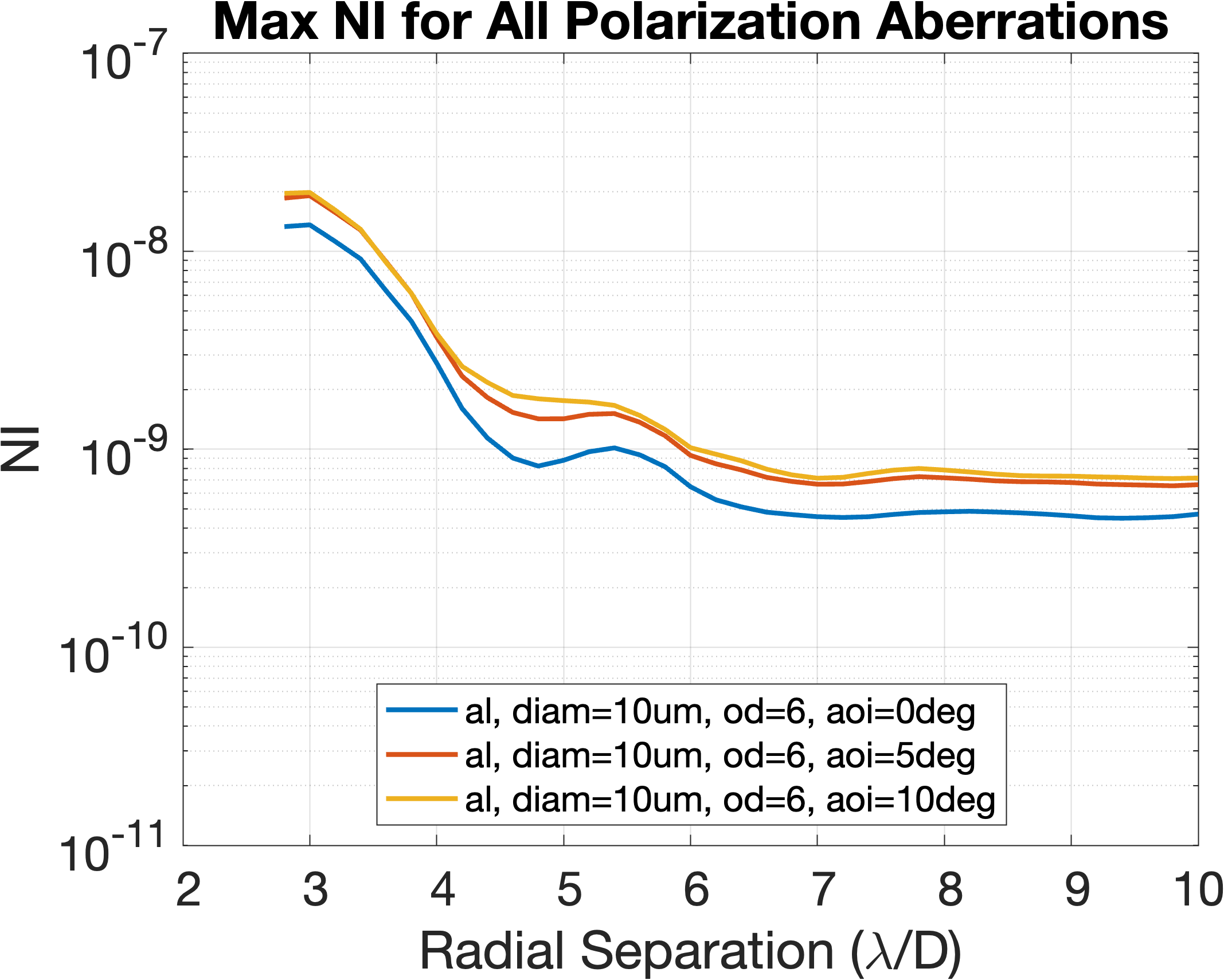}
        \caption{}
        \label{fig:falco_aoi_max_polab}
    \end{subfigure}
    \caption{Radial NI curves for varying AOI on occulting spot. Increasing AOI from $0^{\circ}$ to $10^{\circ}$ improves scalar performance but increases leakage from polarization aberrations, which dominate.}
\end{figure}

The effect of AOI on NI is shown in Figures \ref{fig:falco_aoi_mean_efield}-\ref{fig:falco_aoi_max_polab}.
When considering only the mean electric field, $5^{\circ}$ AOI delivers comparable NI to the normal incidence case up to $6 \lambda/D$ separation, and outperforms normal incidence at large angles.
For $10^{\circ}$ AOI, the performance is improved compared to both normal and $5^{\circ}$ AOI cases for the whole dark hole range.
However, when the full polarization aberrations of the occulter are accounted for, this trend flips, with $10^{\circ}$ AOI performing marginally worse than $5^{\circ}$ AOI, and both raising the leakage significantly over the normal incidence case.

\begin{figure}[ht]
    \centering
    \begin{subfigure}{0.4\linewidth}
        \centering
        \includegraphics[width=\linewidth]{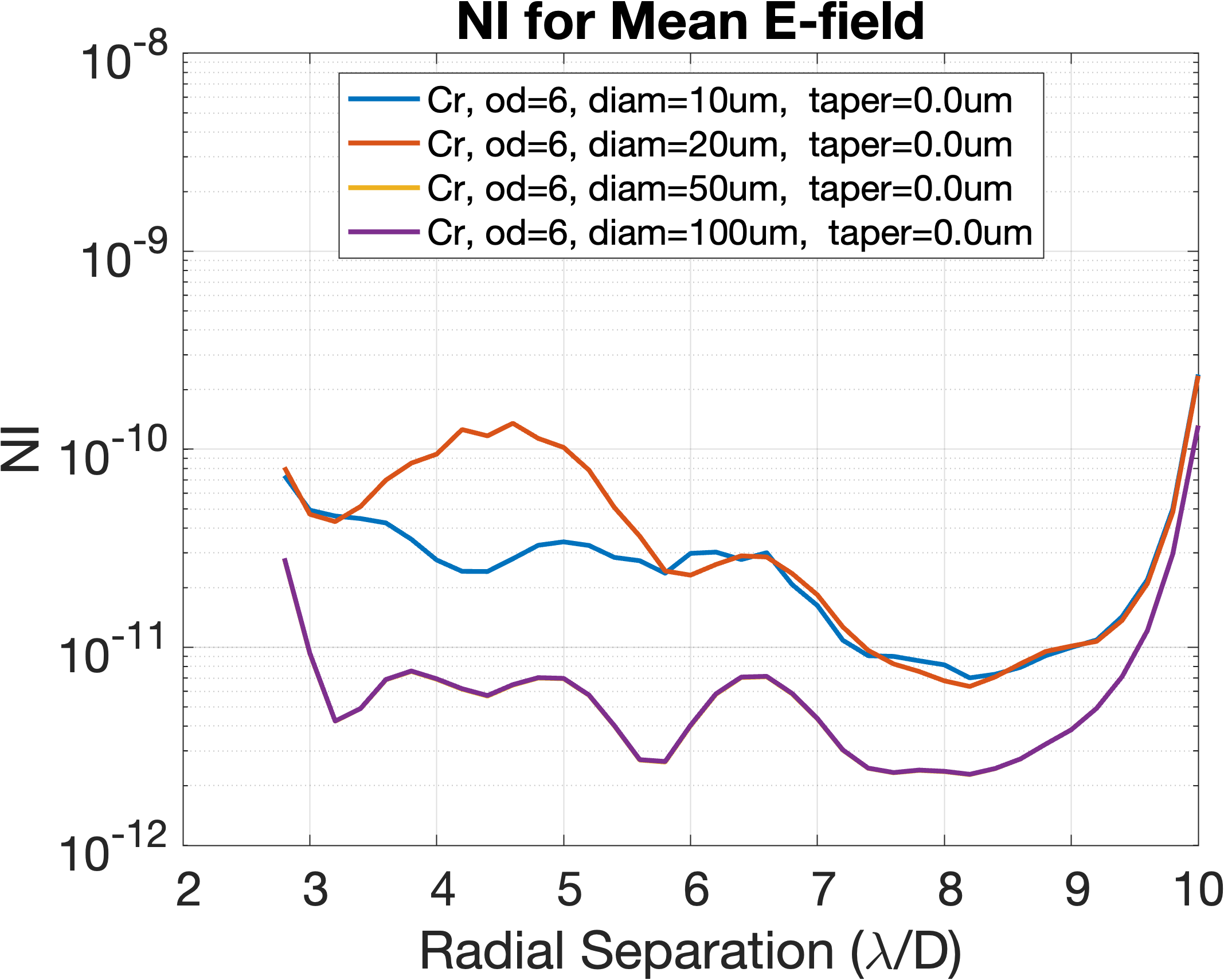}
        \caption{}
        \label{fig:falco_diam_mean_efield}
    \end{subfigure}
    \begin{subfigure}{0.4\linewidth}
        \centering
        \includegraphics[width=\linewidth]{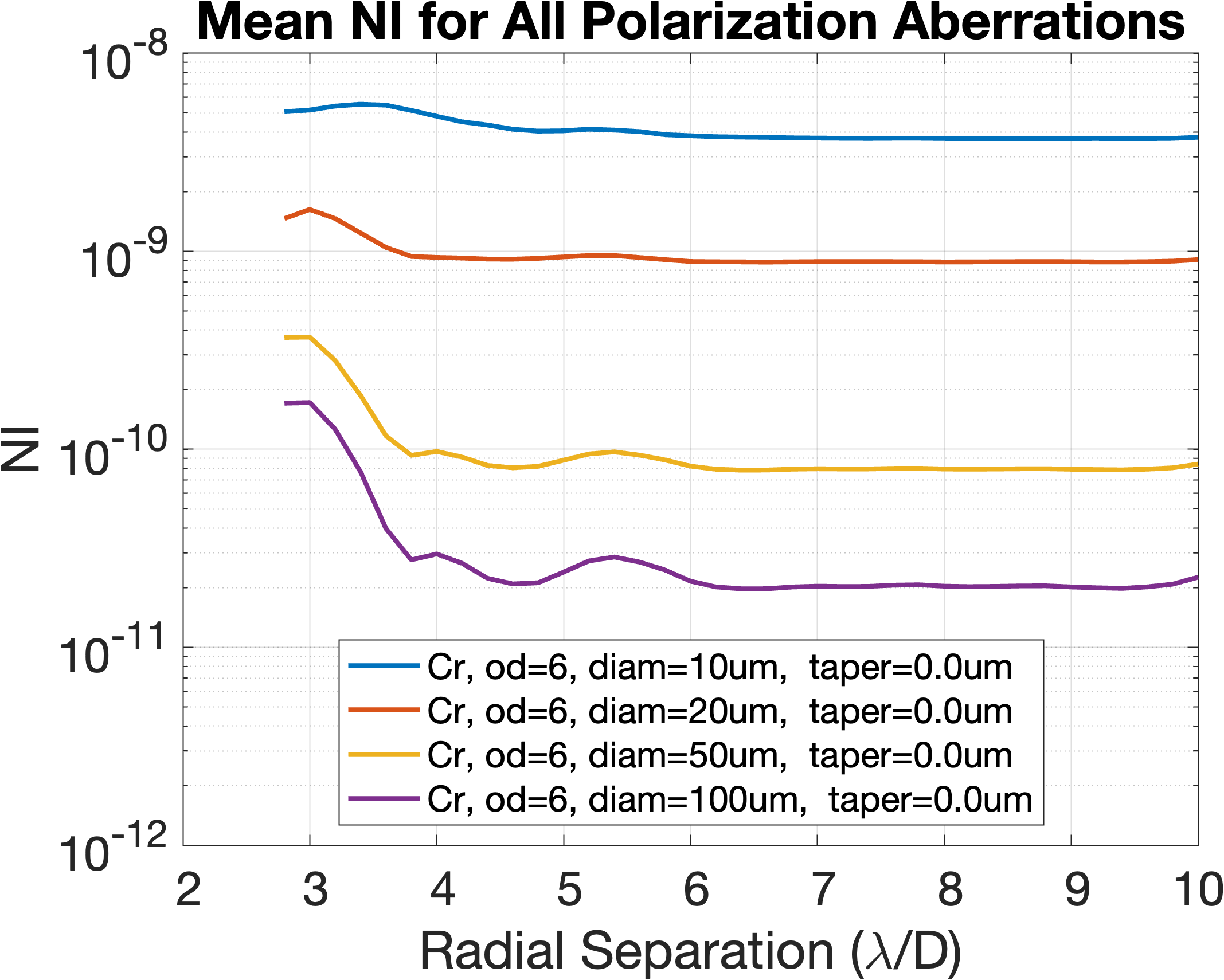}
        \caption{}
        \label{fig:falco_diam_mean_polab}
    \end{subfigure}
    \begin{subfigure}{0.4\linewidth}
        \centering
        \includegraphics[width=\linewidth]{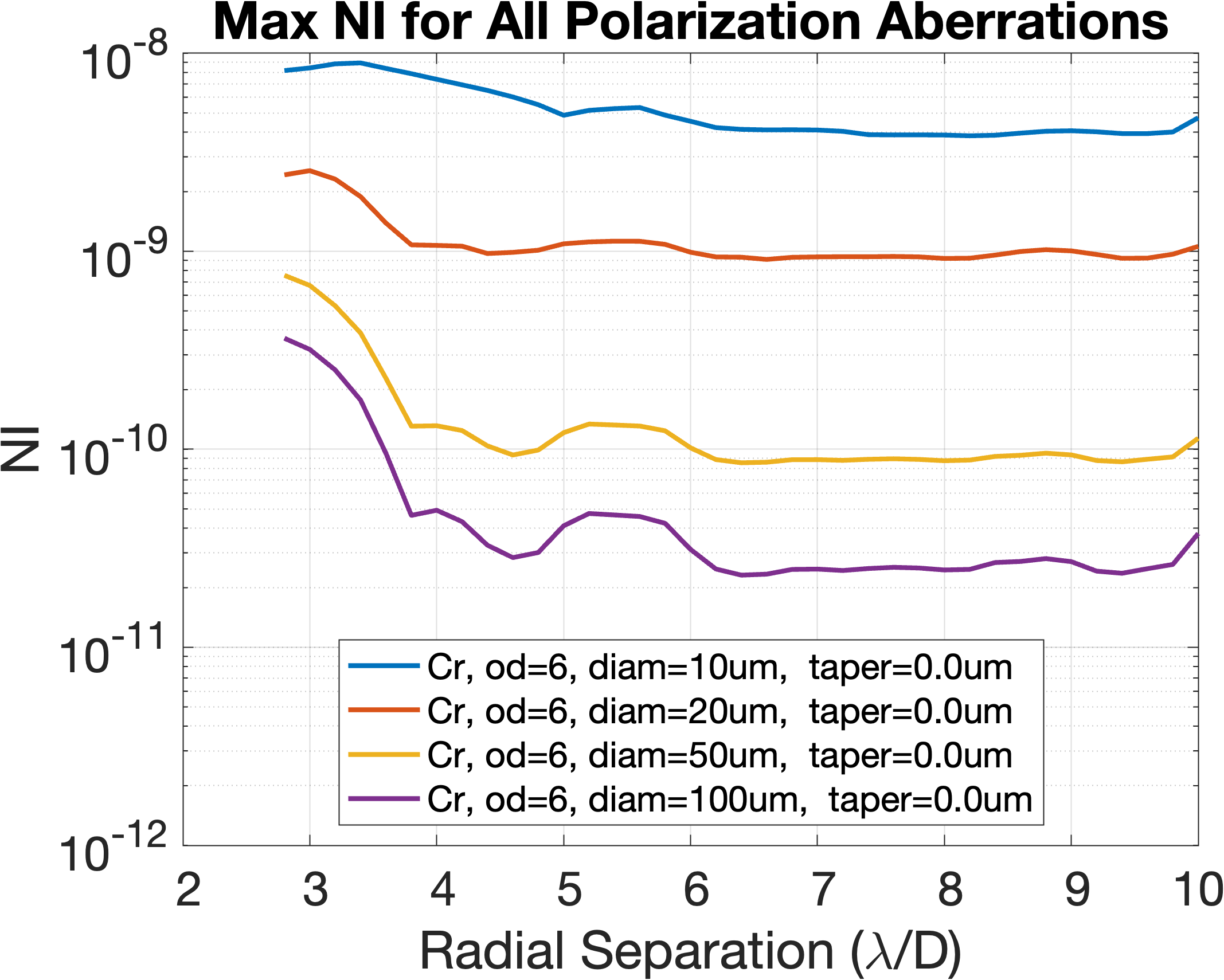}
        \caption{}
        \label{fig:falco_diam_max_polab}
    \end{subfigure}
    \caption{Radial NI curves for varying occulting spot diameters. Leakage from polarization aberrations reduces as diameter increases.}
\end{figure}

Occulting spot diameter has a much larger effect on leakage than AOI.
Mean E-field NI for $10$ and $20$ \si{\micro \meter} diameter cylindrical occulters, shown in Figure \ref{fig:falco_diam_mean_efield}, is below the $10^{-10}$ level, save for a bump around $4.5 \lambda/D$ in the $20$ \si{\micro \meter} case.
The NI curves for $50$ and $100$ \si{\micro \meter} are essentially identical to each other and below $10^{-11}$ between $3-9.5 \lambda/D$.
Including the polarization aberrations, the advantage of larger diameter occulters becomes pronounced.
The $10$ \si{\micro \meter} spot reaches mid to high $10^{-9}$ mean NI and the $20$ \si{\micro \meter} spot reaches bottom $10^{-9}$ mean NI. 
Increasing the spot diameter improves performance considerably, with $50$ \si{\micro \meter} diameter affording $10^{-10}$ mean NI at $4 \lambda/D$ and up, and $100$ \si{\micro \meter} reaching mid $10^{-11}$ mean NI.
Both larger diameters see nearly an order of magnitude jump in leakage from $4 \lambda/D$ down to the $2.8\lambda/D$ mask radius.

\begin{figure}[ht]
    \centering
    \begin{subfigure}{0.4\linewidth}
        \centering
        \includegraphics[width=\linewidth]{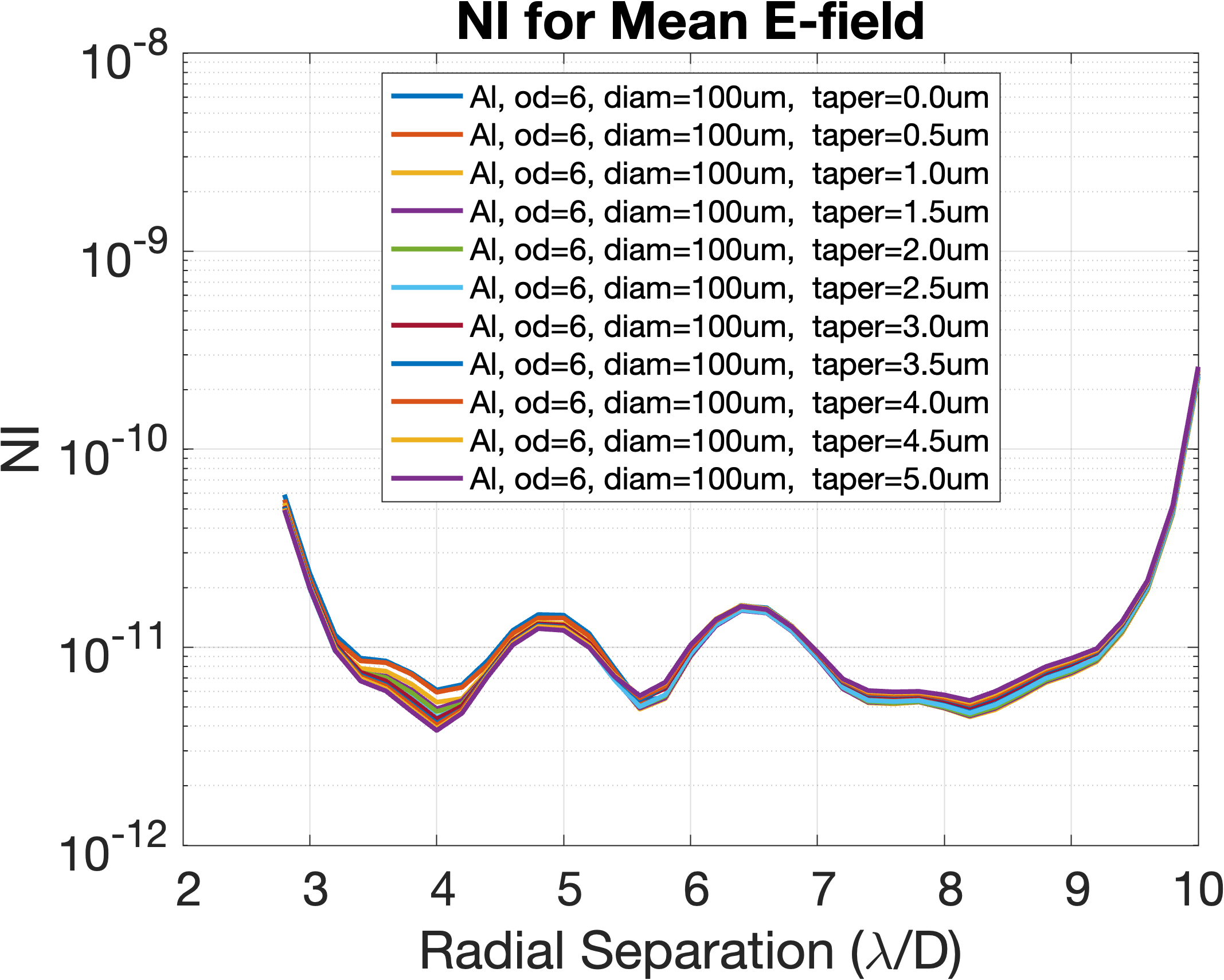}
        \caption{}
        \label{fig:falco_taper_mean_efield}
    \end{subfigure}
    \begin{subfigure}{0.4\linewidth}
        \centering
        \includegraphics[width=\linewidth]{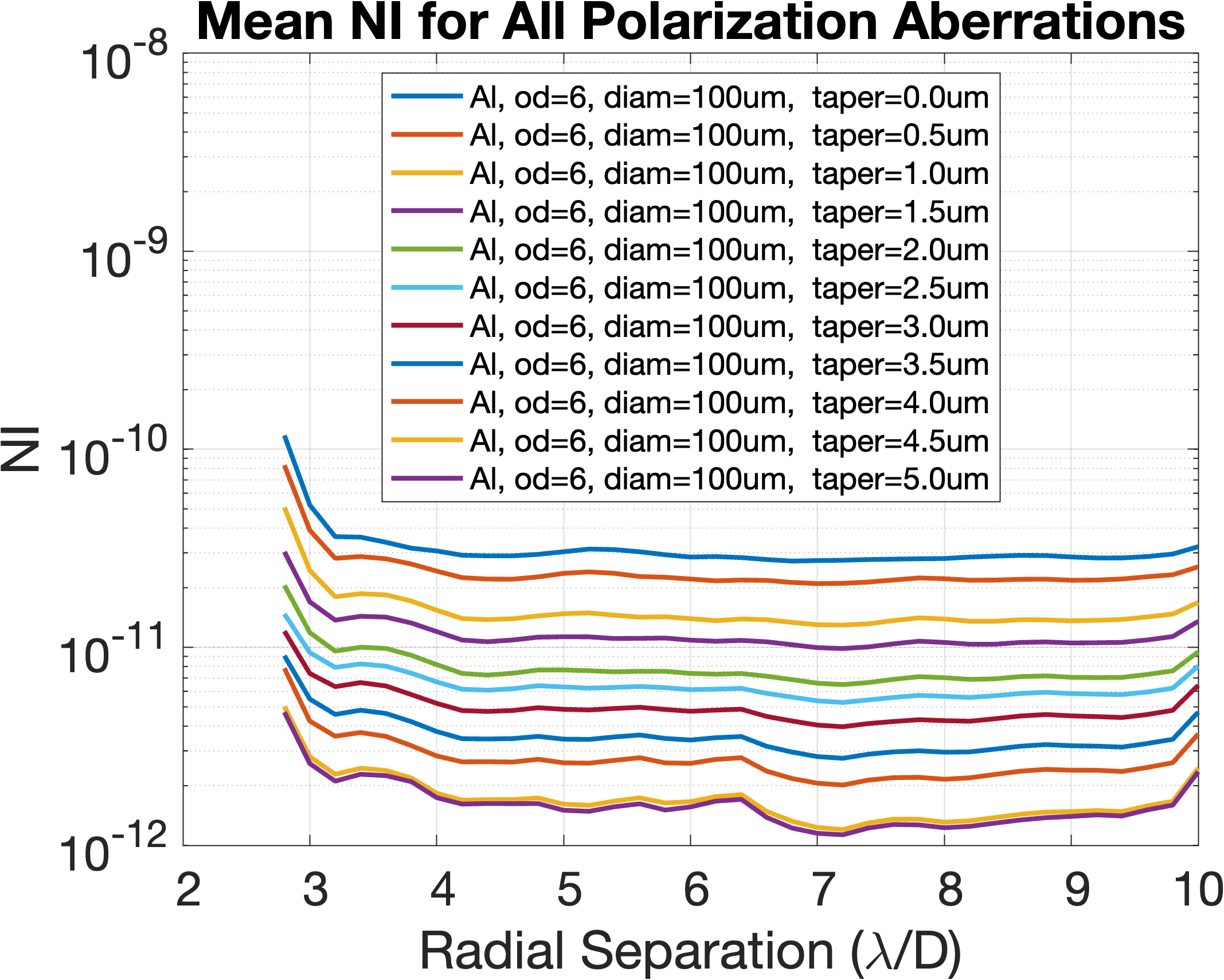}
        \caption{}
        \label{fig:falco_taper_mean_polab}
    \end{subfigure}
    \begin{subfigure}{0.4\linewidth}
        \centering
        \includegraphics[width=\linewidth]{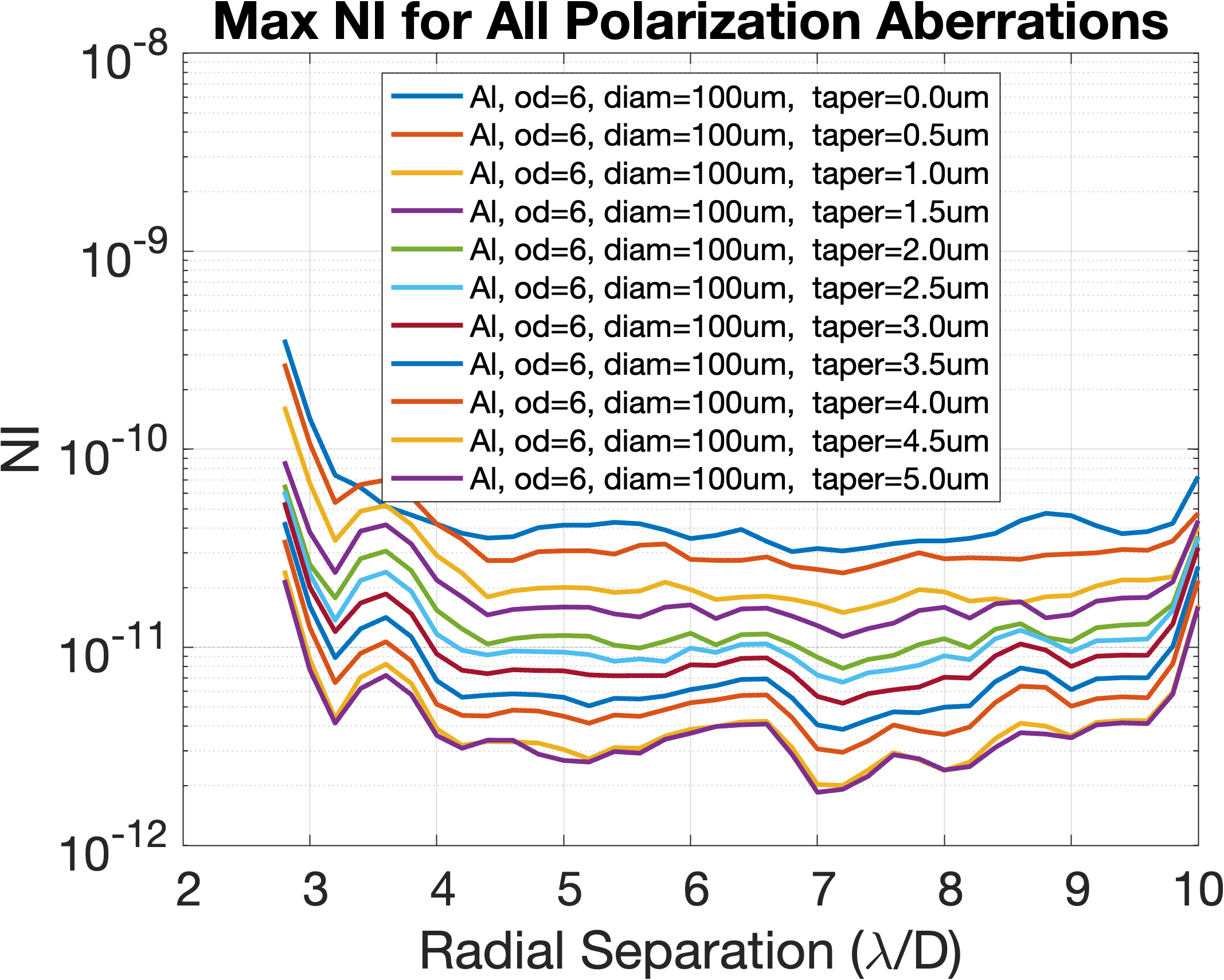}
        \caption{}
        \label{fig:falco_taper_max_polab}
    \end{subfigure}
    \caption{Radial NI curves for an array of occulting spot taper widths. Taper has a minimal effect on leakage due to scalar aberrations, but leakage from polarization aberrations is reduced by $>10$ as taper is increased from $0$ \si{\micro \meter} to $5$ \si{\micro \meter}.}
    \label{fig:falco_taper}
\end{figure}

The width of the transition region has very little effect on NI for the mean electric field, but with polarization aberrations included, the effect of edge smoothing is substantial as shown in Fig.~\ref{fig:falco_taper}.
For an OD $6$ aluminum $100$ \si{\micro \meter} spot, choosing the transition width to be $5$ \si{\micro \meter} improves incoherent NI by greater than an order of magnitude.
Between $0$ and $4.5$ \si{\micro \meter} transition width, each $0.5$ \si{\micro \meter} additional width provides roughly $1$ \si{\decibel} increased starlight suppression.
From $4.5$ to $5$ \si{\micro \meter} transition width, the improvement is marginal.
It is possible that even wider transition widths would continue to show significantly improved incoherent NI, but such a spot would likely be more challenging to fabricate.

\begin{figure}[ht]
    \centering
    \begin{subfigure}{0.4\linewidth}
        \centering
        \includegraphics[width=\linewidth]{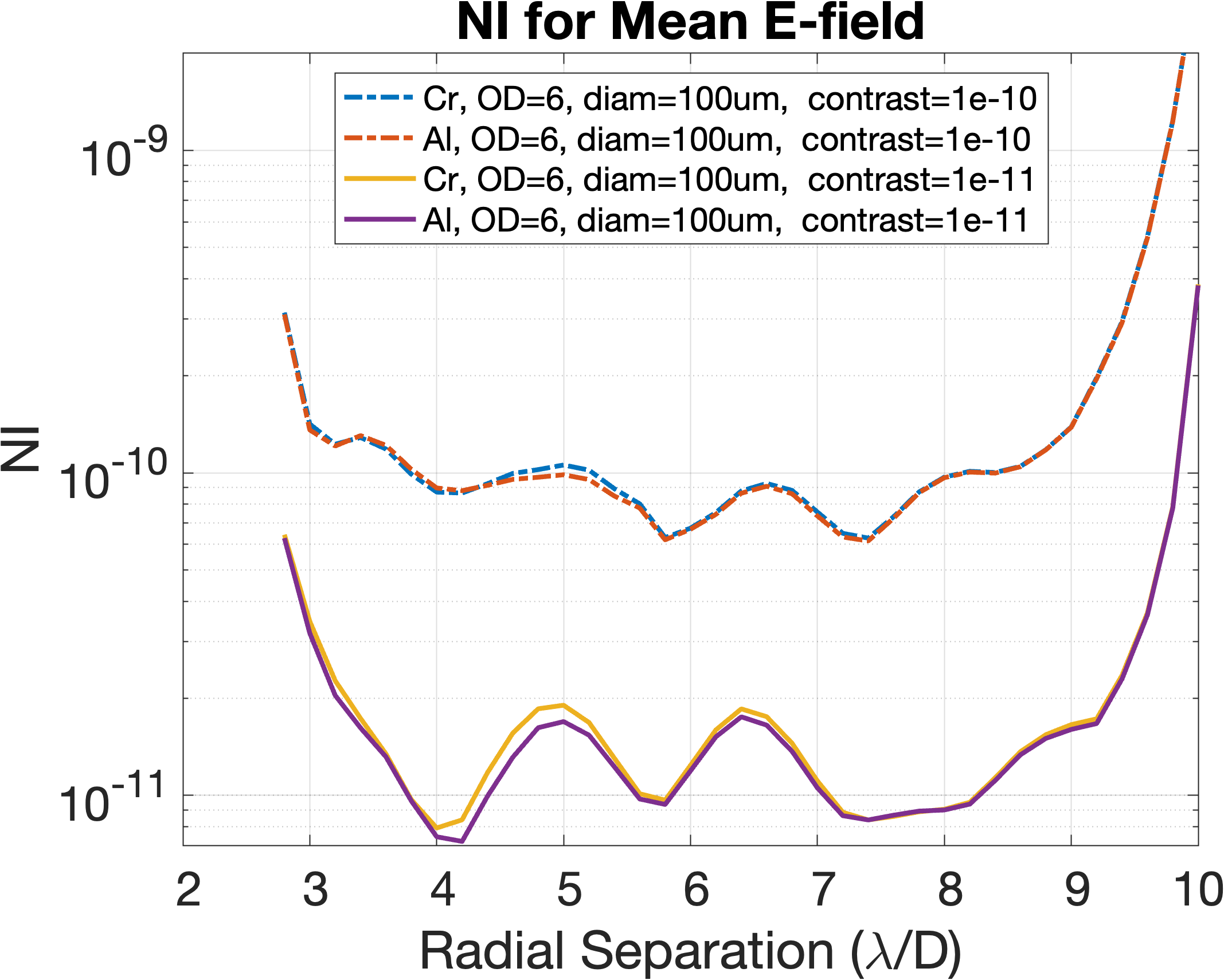}
        \caption{}
        \label{fig:falco_contrast_mean_efield}
    \end{subfigure}
    \begin{subfigure}{0.4\linewidth}
        \centering
        \includegraphics[width=\linewidth]{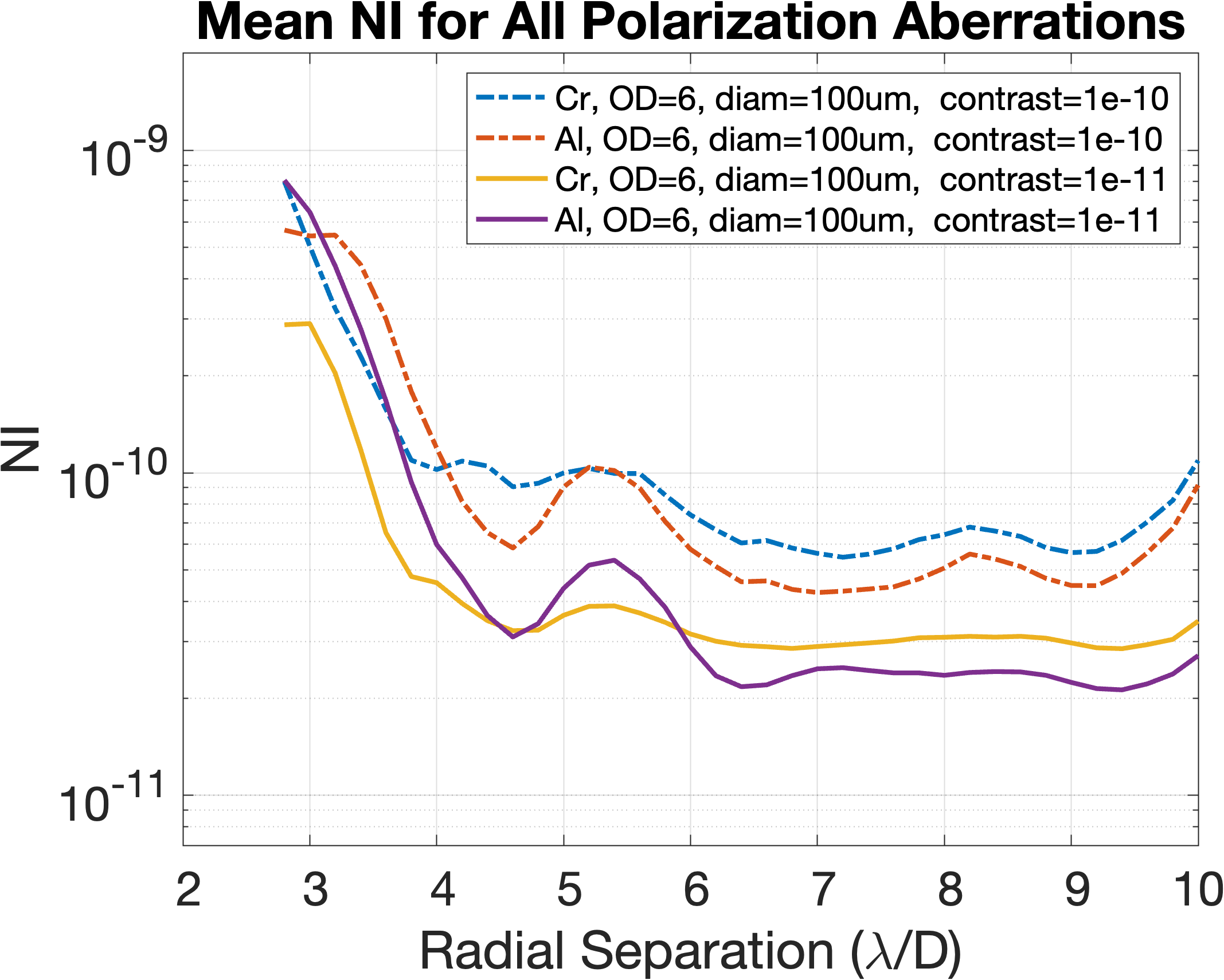}
        \caption{}
        \label{fig:falco_contrast_mean_polab}
    \end{subfigure}
    \begin{subfigure}{0.4\linewidth}
        \centering
        \includegraphics[width=\linewidth]{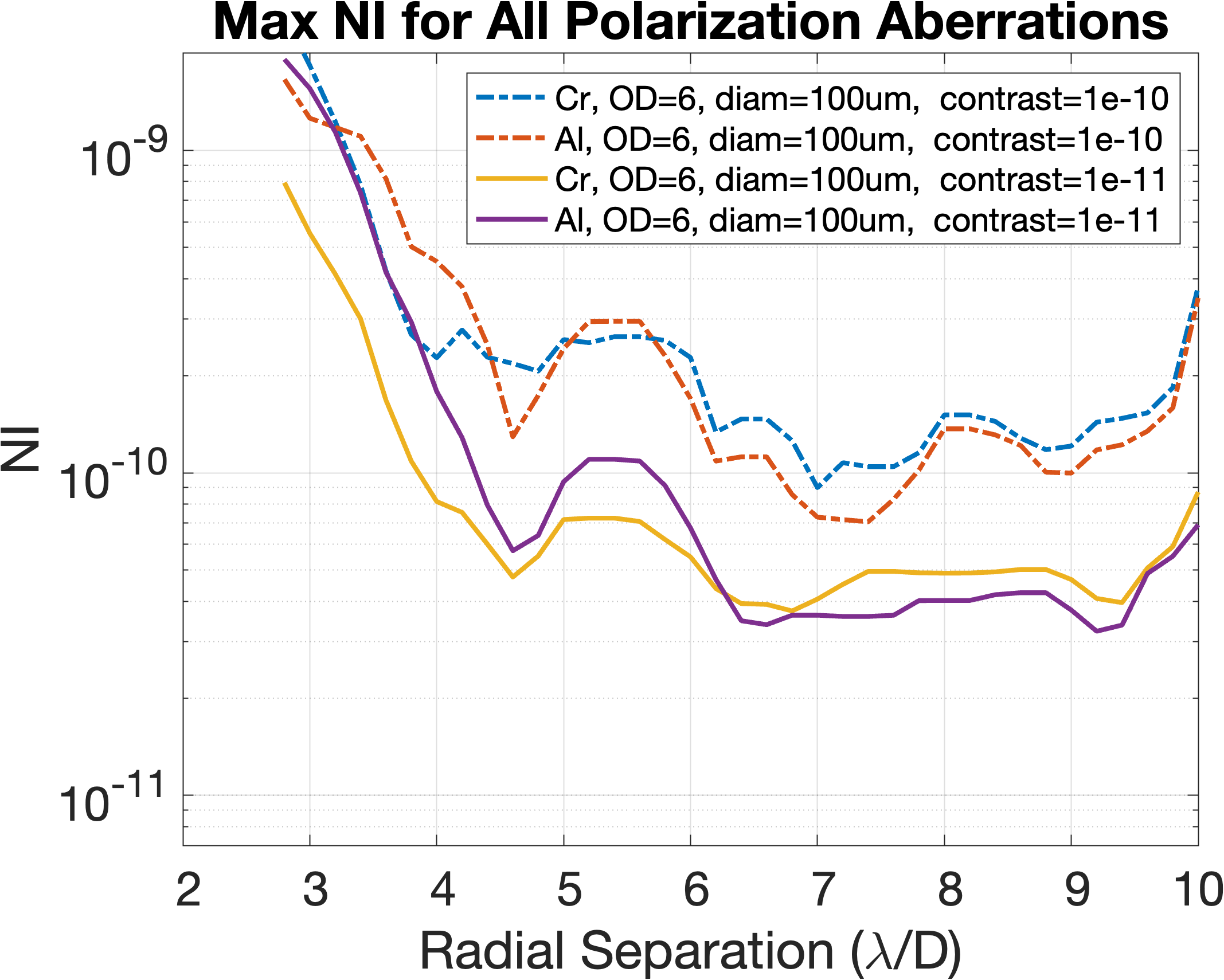}
        \caption{}
        \label{fig:falco_contrast_max_polab}
    \end{subfigure}
    \caption{Radial NI curves for nominally $10^{-10}$ and $10^{-11}$ NI dark holes dug with EFC on the mean electric field.}
    \label{fig:falco_contrast}
\end{figure}

To investigate how leakage from polarization aberrations scales with nominal dark hole NI, dark holes were dug with EFC on the mean electric field down to the $10^{-10}$ and $10^{-11}$ levels. Results are shown in Fig.~\ref{fig:falco_contrast}. Polarization aberrations lead to increased maximum NI in both cases, but the maximum NI is higher for the $10^{-10}$ mean electric field dark hole solution. This indicates that a deeper scalar dark hole can limit the effect of polarization aberrations.

\section{Summary and Future Work}
Focal plane occulter masks create significant polarization aberrations at exo-Earth contrasts.
In order to get $<10^{-10}$ total raw contrast at $\sim 3 \lambda/D$, the mean E-field has to be below $10^{-10}$, and darker mean E-field dark holes limit leakage from polarization effects.
Leakages can be reduced by increasing the spot diameter (larger $\text{F}/\#$), with $\sim 100$ \si{\micro \meter} being the smallest diameter to provide $<10^{-10}$ leakage without edge smoothing.
Smoothing the edge of the spot over a $5$ \si{\micro \meter} transition width reduces leakage by greater than $1$ order of magnitude.
This may enable the use of smaller occulters at lower $\text{F}/\#$ while still achieving exo-Earth contrasts for the HWO, allowing for tighter packaging.

Future simulations will include a vortex coronagraph with occulting dots over the singularity, another candidate for the HWO coronagraph instrument.
Additionally, since the manufacturability of the tapered spots simulated in this work is not well known, prototypes of occulting spots with engineered edge profiles should be fabricated to develop a process.
This would allow testbed results comparing cylindrical and tapered spots.

\acknowledgments 
 
The research was carried out in part at the Jet Propulsion Laboratory, California Institute of Technology, under a contract with the National Aeronautics and Space Administration (80NM0018D0004).

\bibliography{report} 
\bibliographystyle{spiebib} 

\end{document}